\documentclass[
    reprint,
    amsmath,
    amssymb,
    aps,letterpaper,accepted=2022-01-30
]{quantumarticle}
\pdfoutput=1

\usepackage{graphicx}
\graphicspath{ {images/} }
\usepackage{bm}
\usepackage{floatrow}
\usepackage{braket}

\usepackage{listings}
\usepackage{hyperref}
\usepackage{comment}
\usepackage{bbm}
\usepackage{xcolor}
\usepackage{caption}
\usepackage[numbers]{natbib}
\usepackage{notoccite}

\usepackage{amsmath}
\usepackage{amssymb}
\usepackage{amsthm}

\theoremstyle{definition}

\begin{document}

\title{Noise-resistant Landau-Zener sweeps from geometric curves}

\author{Fei Zhuang}
\affiliation{Department of Physics, Virginia Tech, Blacksburg, Virginia 24061, USA}
\author{Junkai Zeng}
\affiliation{Department of Physics, Virginia Tech, Blacksburg, Virginia 24061, USA}
\affiliation{Shenzhen Institute of Quantum Science and Engineering,
Southern University of Science and Technology, Shenzhen, Guangdong 518055, China}
\author{Sophia E. Economou}
\affiliation{Department of Physics, Virginia Tech, Blacksburg, Virginia 24061, USA}
\author{Edwin Barnes}
\affiliation{Department of Physics, Virginia Tech, Blacksburg, Virginia 24061, USA}
\email{efbarnes@vt.edu}

\begin{abstract}
Landau-Zener physics is often exploited to generate quantum logic gates and to perform state initialization and readout. The quality of these operations can be degraded by noise fluctuations in the energy gap at the avoided crossing. We leverage a recently discovered correspondence between qubit evolution and space curves in three dimensions to design noise-robust Landau-Zener sweeps through an avoided crossing. In the case where the avoided crossing is purely noise-induced, we prove that operations based on monotonic sweeps cannot be robust to noise. Hence, we design families of phase gates based on non-monotonic drives that are error-robust up to second order. In the general case where there is an avoided crossing even in the absence of noise, we present a general technique for designing robust driving protocols that takes advantage of a relationship between the Landau-Zener problem and space curves of constant torsion.
\end{abstract}

\maketitle


\section{Introduction}\label{sec:intro}
Landau-Zener (LZ) transitions have been studied extensively since the beginning of quantum mechanics \cite{landau1932,stuckelberg1932,zener1932non,majorana1932}. Such transitions occur when a system is tuned through an avoided crossing. In the context of quantum information processing, LZ transitions are widely used to perform qubit initialization, control, and readout and to characterize the system Hamiltonian \cite{sun2009population,petta2010coherent,diCarlo2009demonstration,DiCarlo2010preparation,sun2010tunable,Mariantoni2011Implementing,ReedNature2012,cao2013ultrafast,thiele2014electrically,martinis2014fast,wang2018landau,rol2019fast}. Many of these applications are carried out via Landau-Zener-Stuckelberg interferometry, which is analogous to Mach-Zehnder interferometry~\cite{shevchenko2010landau,shytov2003landau,ji2003electronic}. This has been employed for example in quantum dots, superconducting circuits, and donors in nanowires~\cite{petta2010coherent, sillanpaa2006continuous,oliver2005mach,dupont2013coherent,nalbach2013nonequilibrium}.

\begin{figure}
    \centering
    \includegraphics[width=0.8\columnwidth]{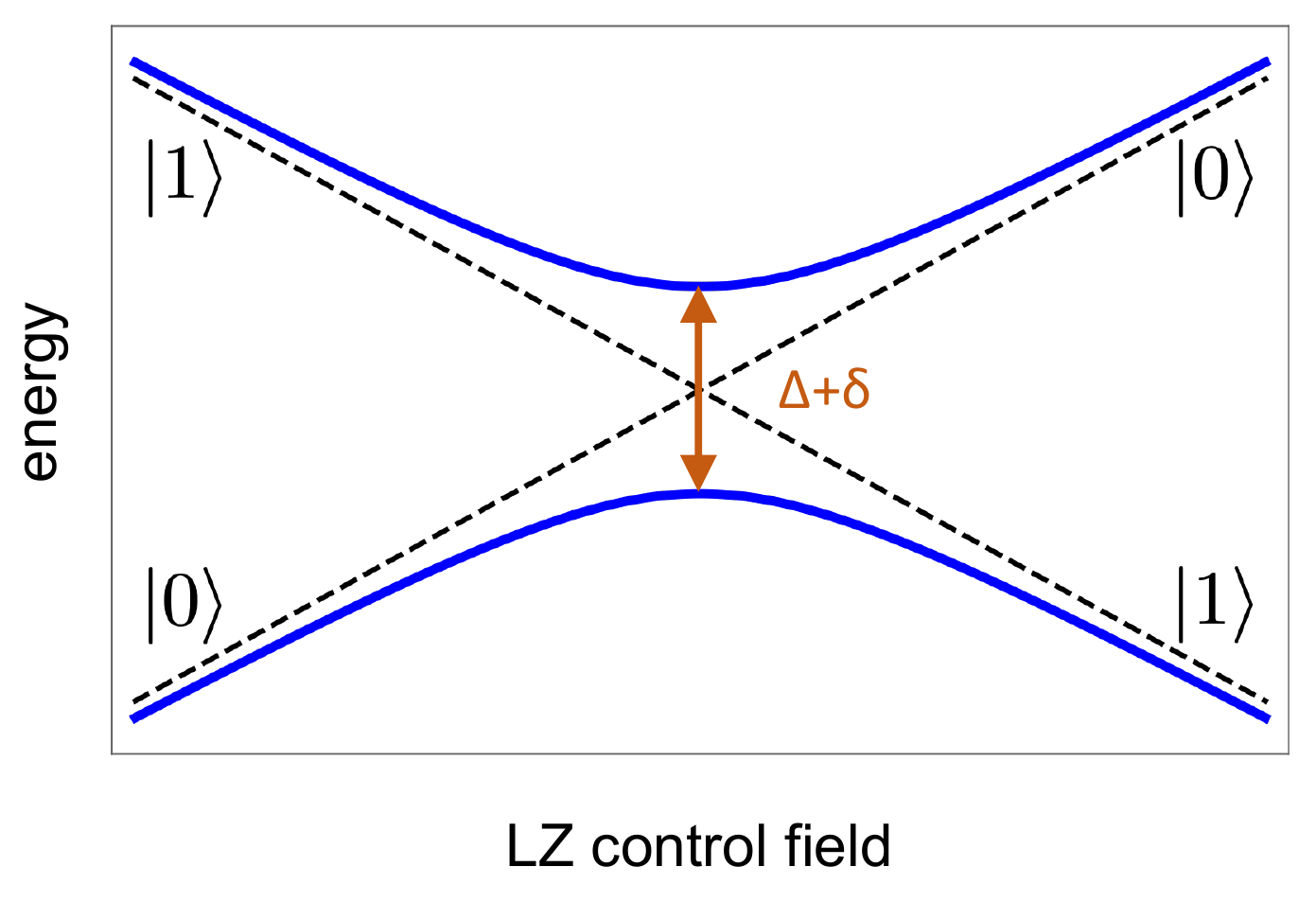}
    \caption{Noisy avoided crossing in which the energy gap $\Delta$ is stochastically shifted by an amount $\delta\ll\Delta$. If the gap is zero in the noiseless case ($\Delta=0$), the noise can still open a gap to create an avoided crossing, causing unwanted transitions.}
    \label{fig:LZschematic}
\end{figure}

The performance of LZ sweeps is highly susceptible to noise and decoherence~\cite{huang2011landau}. This can be particularly problematic both for sweeping through a level crossing and through an avoided crossing. In the former case, noise can open a small gap (see Fig.~\ref{fig:LZschematic}), rendering what is nominally a crossing into an anticrossing and thus causing an unwanted transition with some finite probability. Such a scenario can occur for example in multi-qubit superconducting processors where two-qubit gates are implemented via tuning (DC driving)~\cite{diCarlo2009demonstration,DiCarlo2010preparation,sun2010tunable,Mariantoni2011Implementing} rather than with microwaves (AC driving). Similarly, if the goal is to bring the system to a particular level on the far side of an avoided crossing, noise can cause unwanted transitions to the other level. In cases where avoided crossings are used to implement quantum operations or perform initialization and readout, noise can reduce the performance. Several general methods have been developed to dynamically suppress environmental noise and improve qubit operation fidelities~\cite{Viola1998dynamical,viola1999dynamical,viola1999universal,viola2003robust,Brown2004arbitrarily,khodjasteh2005fault,uhrig2007keeping,Uys2009optimized,west2010high,Biercuk2011dynamical,Wang2012composite,khodjasteh2013designing,CalderonVargas2017dynamically}. In the particular case of LZ control, this issue has been addressed through the use of composite LZ pulses~\cite{hicke2006fault}, super-adiabatic LZ pulses~\cite{bason2012high, martinis2014fast}, and geometric-phase LZ sweeps~\cite{gasparinetti2011geometric,tan2014demonstration,zhang2014realization,wang2016experimental}. However, these approaches can involve long control times, experimentally challenging control waveforms, or imperfect noise cancellation. 

In this paper, we present a general method for designing noise-resistant LZ sweeps through avoided crossings. We do this by leveraging a recently developed geometric formalism for designing noise-resistant quantum control called Space Curve Quantum Control (SCQC) \cite{barnes2021dynamically,zeng2018general,zeng2018fastest,zeng2019geometric}. This approach utilizes a surprising connection between the evolution of a qubit and geometric curves in two or three dimensions. Through this connection, we show that the classic LZ problem, where the system is driven at constant velocity through an avoided crossing, can be mapped to an Euler spiral (see Fig.~\ref{fig:LZ cornu}). This is a well-known curve with the key feature that its curvature changes linearly; Euler spirals are used in diverse fields ranging from railroad engineering to integrated photonic circuits~\cite{Levien:EECS-2008-111,Bartholdi2012,MatteoOSA2013,LiLSA2018}. Interestingly, it has been shown that peeling an orange into a constant-width strip produces an Euler spiral in the zero-width limit when the strip is laid out on a flat surface~\cite{Bartholdi2012}. 
Using the connection between LZ physics and Euler spirals, we prove that there is no monotonic LZ driving protocol that can bring a system through a noise-induced anticrossing without accumulating errors at first order in the noise. This guides us to construct non-monotonic sweeps that suppress noise. In addition, we design arbitrary phase gates that are noise-resistant. We also show how noise can be cancelled when the gap at the avoided crossing fluctuates about a nonzero value. Robust LZ sweeps in this case correspond to closed curves of constant torsion in three dimensions. We present a general method for obtaining such curves, and we apply this to design explicit LZ sweep protocols that cancel noise to first order. We confirm the robustness of our protocols by numerically computing the operation fidelity as a function of noise strength. 

The paper is organized as follows. We
begin by reviewing the LZ problem and the SCQC formalism in Sec.~\ref{sec:GFreview}. In Sec.~\ref{sec:zerogap}, we use SCQC to design robust LZ pulses in the case where the avoided crossing has no energy gap in the absence of noise. We then show how noise-resistant LZ sweeps can be obtained from closed curves of constant torsion for nonzero gaps in Sec.~\ref{sec:nonzerogap}. We conclude in Sec.~\ref{sec:conclusion}. Several appendices contain further examples and mathematical details about our approach, including phase gate designs using a noisy crossing. The appendices also contain the proof that it is not possible to cancel noise by sweeping through an avoided crossing monotonically.
\section{Landau-Zener transitions and Geometric Formalism}\label{sec:GFreview}

The original LZ problem concerns a two-level system that is dynamically tuned through an avoided crossing \cite{landau1932,stuckelberg1932,zener1932non,majorana1932}. This can be described by the Hamiltonian
\begin{equation}\label{eq:LZham}
H_0(t)=\frac{\Omega(t)}{2}\sigma_z+\frac{\Delta}{2}\sigma_x,
\end{equation}
where $\Omega(t)$ is a detuning parameter or energy bias, while $\Delta$ is the (constant) energy gap at zero bias. $\sigma_z$ and $\sigma_x$ are Pauli matrices, and we label the basis states as $\ket{0}$ and $\ket{1}$. These states are often referred to as the diabatic states. The instantaneous eigenstates (i.e., the adiabatic states) of $H_{0}$ form an avoided crossing at the zero bias point, $\Omega=0$ (see Fig.~\ref{fig:LZschematic}). When $|\Omega|\gg\Delta$, the ground state approaches $\ket{1}$ when $\Omega>0$, and it approaches $\ket{0}$ when $\Omega<0$. The original works \cite{landau1932,stuckelberg1932,zener1932non,majorana1932} focused on a linear sweep, $\Omega(t)=vt$, and computed the probability that the system remains in the same diabatic state it started in (or equivalently that a transition between adiabatic states occurs) as a function of the sweep velocity $v$, resulting in the famous formula $P_{\ket{0}\rightarrow\ket{0}}=e^{-\pi\Delta^2/2v}$. It follows that the system remains in state $\ket{0}$ on the far side of the avoided crossing when $\Delta$ is small and/or $v$ is large. Thus, when $\Omega$ is quickly tuned from large negative values to large positive values, there is a high probability that the system will remain in its initial diabatic state at the end of the $\Omega$ sweep. However, when $\Omega$ is tuned through the avoided crossing more slowly, there is a finite transition probability. In general, this transition probability depends on the form of $\Omega(t)$. While the linear-sweep case can be solved exactly in terms of parabolic cylinder functions \cite{shevchenko2010landau}, exact analytical solutions cannot be obtained for general choices of $\Omega(t)$ \cite{barnes2012analytically,barnes2013analytically}.

Here, we are interested in the case where the energy gap $\Delta$ fluctuates slowly in time as a consequence of environmental noise. Such fluctuations are common in solid-state systems such as superconducting qubits or semiconductor spin qubits \cite{Krantz2019quantum,Hanson2007spins}. If the noise is sufficiently slow, then it can be treated using the quasistatic approximation in which the fluctuation $\delta$ in the energy gap is constant (but unknown) during the LZ sweep \cite{Krantz2019quantum,martins2016noise}. The total Hamiltonian can then be written as
\begin{equation}\label{eq:totalH}
    H(t)=H_0(t)+\delta H
         =\frac{\Omega(t)}{2} \sigma_z+\frac{\Delta}{2} \sigma_x+\delta\sigma_x.
\end{equation}
\textcolor{black}{When averaged over a distribution of width $\sigma_\delta$, the noise fluctuation $\delta$ gives rise to dephasing on a timescale $T_2^*\sim1/\sigma_\delta$.} Our goal is to remove the effect of $\delta$ by choosing $\Omega(t)$ appropriately. If $\delta$ is sufficiently small, then we can perform a perturbative expansion of the evolution operator at the final time $T$: $U(T)=U_0(T)+\delta U_1(T)+\delta^2U_2(T)+\ldots$ Here, $U_0$ is the noiseless evolution operator generated by $H_0$. Because, $U_1$, $U_2$, $...$ depend on $\Omega(t)$, we can engineer this function so that at least the first few orders vanish at the final time: $U_1(T)=0$, $U_2(T)=0$, ... This is equivalent to making the evolution operator in the interaction picture, $U_I(T)=U_0^\dagger(T)U(T)$, as close to the identity as possible. This in turn can be facilitated by using a Magnus expansion, $U_I(T)=e^{-i(\delta A_1(T) + \delta^2 A_2(T) + O(\delta^3)) }$, where 
\begin{align}
    A_1(t)&= \int_0^t  U_0^\dagger(t_1) \sigma_x U_0(t_1) d t_1, \label{eq:A1} \\
    A_2(t)&= \frac{1}{2}\int_0^t [\dot{A}_1(t_1), A_1(t_1)] dt_1. \label{eq:A2}
\end{align}

Because it is a traceless Hermitian matrix, ${A}_1(t)$ can be parameterized as 
\begin{equation}\label{eq:dot A}
    {A}_1(t)=\mathbf{r}(t) \cdot \mathbf{\sigma},
\end{equation}
where $\sigma$ is the vector of Pauli matrices. It follows from Eq.~\eqref{eq:A1} that $\|\dot{\mathbf{r}}(t)\|=1$, where $\dot f=df/dt$. Thus $\mathbf{r}(t)$ can be interpreted as a curve in \textcolor{black}{$\mathbbm{R}^3$} for which time is the arc length. This is the starting point of the SCQC formalism~\cite{barnes2021dynamically}. Achieving first-order noise robustness requires $A_1(T)=0\Rightarrow\mathbf{r}(T)=0$, or equivalently, the curve $\mathbf{r}$ must be closed. Since $\|\mathbf{r}\|$ provides a measure of how much the evolution deviates from the ideal identity operation, we refer to $\mathbf{r}$ as the {\it error} curve. If we also wish to cancel second-order noise errors, then we additionally need to satisfy $A_2(T)=0$. It follows from Eqs.~\eqref{eq:A2} and \eqref{eq:dot A} that
\begin{equation}\label{A2 parametric}
    A_2(t)=-i \left(\int_0^t \mathbf{r}(t_1)\times\dot{\mathbf{r}}(t_1) dt_1\right)\cdot\sigma.
\end{equation}
Since $\hat a\cdot\int \mathbf{r}\times d\mathbf{r}$ is proportional to the area enclosed by the curve $\mathbf{r}$ after it is projected onto the plane orthogonal to $\hat a$, it follows that second-order noise cancellation is tantamount to requiring the enclosed area to vanish for all three projections, $\hat a=\hat x$, $\hat y$, $\hat z$  \cite{zeng2018general,zeng2019geometric}.

At each point along a curve in \textcolor{black}{$\mathbbm{R}^3$}, one can define an orthonormal frame consisting of the tangent vector $\dot{\mathbf{r}}$, the normal vector $\mathbf{n}=\ddot{\mathbf{r}}/\kappa$, and the binormal vector $\mathbf{b}=\dot{\mathbf{r}}\times\mathbf{n}$. The Frenet-Serret equations, 
\begin{equation}\label{eq:FSframe}
    \begin{pmatrix}
     \ddot{\mathbf{r}}\\
     \dot{\mathbf{n}}\\
     \dot{\mathbf{b}}
    \end{pmatrix}=\begin{pmatrix}
                  0& \kappa &0 \\
                 -\kappa & 0 & \tau\\
                  0&-\tau&0
                    \end{pmatrix} 
     \begin{pmatrix}
     \dot{\mathbf{r}}\\
     \mathbf{n}\\
     \mathbf{b}
    \end{pmatrix},
\end{equation}
describe how these three vectors evolve along the curve. The curvature $\kappa(t)$ describes how quickly the tangent vector $\dot{\mathbf{r}}(t)$ is changing direction at a given point $t$, while the torsion $\tau$ describes the rate at which the curve is bending out of the plane spanned by $\dot{\mathbf{r}}(t)$ and $\mathbf{n}(t)$ at point $t$. It is important to note that each of the vectors $\dot{\mathbf{r}}(t)$, $\mathbf{n}(t)$, $\mathbf{b}(t)$ can be interpreted as a curve in its own right and thus provides an alternative way to view the evolution geometrically. These complementary descriptions can serve as powerful tools in constructing curves obeying constraints, as we show in the following sections where we obtain explicit examples of robust LZ sweeps.

Once we construct a closed curve (perhaps with vanishing-area projections as well), we can extract noise-cancelling driving fields from it. This is achieved by relating $\kappa$ and $\tau$ to $H_0$ \cite{zeng2019geometric,Lehto2015geometry}. From Eqs.~\eqref{eq:A1} and \eqref{eq:dot A}, we have $\dot{A}_1=\dot{\mathbf{r}}\cdot\sigma$ and $\ddot{A}_1=-\Omega U_0^\dag\sigma_yU_0=\ddot{\mathbf{r}}\cdot\sigma=\kappa\mathbf{n}\cdot\sigma$. We can therefore define the Frenet-Serret frame as
\begin{equation}
    \dot{\mathbf{r}}\cdot\sigma=U_0^\dag\sigma_xU_0,\quad
    \mathbf{n}\cdot\sigma=U_0^\dag\sigma_yU_0,\quad
    \mathbf{b}\cdot\sigma=U_0^\dag\sigma_zU_0.
\end{equation}
From these expressions, we find that the LZ pulse and energy gap are given by (up to a sign) the curvature and torsion:
\begin{align}\label{eq:curvature}
    \kappa=\ddot{\mathbf{r}}\cdot\mathbf{n}&=\frac{1}{2}\hbox{Tr}\left\{(\ddot{\mathbf{r}}\cdot\sigma)(\mathbf{n}\cdot\sigma)\right\}\nonumber\\&=\frac{i}{2}\hbox{Tr}\left\{[H_0,\sigma_x]\sigma_y\right\}=-\Omega,
\end{align}
\begin{align}\label{eq:torsion}
    \tau=\dot{\mathbf{n}}\cdot\mathbf{b}&=\frac{1}{2}\hbox{Tr}\left\{(\dot{\mathbf{n}}\cdot\sigma)(\mathbf{b}\cdot\sigma)\right\}\nonumber\\&=\frac{i}{2}\hbox{Tr}\left\{[H_0,\sigma_y]\sigma_z\right\}=-\Delta.
\end{align}
Thus, we see that the fields that implement a cancellation of leading-order quasistatic noise errors can be obtained by constructing closed curves and computing their curvature and torsion. The SCQC formalism shines light on how noise-robustness properties depend on the pulse shape. Moreover, it imposes the minimal constraints necessary to ensure error cancellation. The method is in fact completely general: Any pulse that cancels noise to leading order corresponds to a closed curve. The cancellation of higher-order noise errors can be accomplished by further imposing certain vanishing-volume constraints on the curve~\cite{zeng2018general}. \textcolor{black}{In Ref.~\cite{zeng2019geometric}, it was shown that arbitrary, noise-resistant single-qubit gates can be obtained using this method.} The SCQC formalism has been further extended to multi-level systems and to multiple noise sources and time-dependent noise \cite{buterakos2020geometrical,barnes2015robust,gungordu2019analytically,throckmorton2019conditions,dong2021doubly,li2021designing}. \textcolor{black}{We note in passing that other geometric reverse-engineering approaches to error-correcting control have also been devised based on alternative parameterizations of the evolution operator~\cite{dridi2020optimal,daems2013robust}. In this work, we focus on SCQC because of the simple geometrical interpretation of noise cancellation (closed curve) it affords and because of the direct connection between torsion and the energy gap $\Delta$ (Eq.~\eqref{eq:torsion}). In general, the curvature and torsion of a curve in $\mathbbm{R}^3$ both vary along the curve. For our LZ problem, we are interested in the case of a constant energy gap $\Delta$, which according to Eq.~\eqref{eq:torsion} means that we need to focus on constant-torsion curves. Thus, to construct LZ sweeps to first order, we must find closed curves of constant torsion. When $\Delta\ne0$, this is a nontrivial task \cite{weiner1974closed,weiner1977closed} that has not been addressed in any of the previous works on the SCQC formalism. We present a general solution to this problem in Sec.~\ref{sec:nonzerogap}. The ability to construct closed curves of constant torsion is critical not only to the LZ problem, but also to any context in which a qubit is driven by a pulse with constant detuning. This general Rabi problem arises throughout quantum physics.}

At this point it is worth noting a subtle difference between the geometric formalism introduced in Ref.~\cite{zeng2019geometric} and the one described here. Usually, the curvature of a curve in \textcolor{black}{$\mathbbm{R}^3$} is defined to be $\|\ddot{\mathbf{r}}\|$, which is strictly nonnegative. This definition would in turn force $\Omega(t)$ to always have the same sign, which was the case in Ref.~\cite{zeng2019geometric}. Here, we generalize the definition of the curvature slightly to allow it to assume both positive and negative values during the evolution. This allows us to construct robust LZ sweeps that pass through the avoided crossing at $\Omega=0$. The sign of the curvature is chosen such that $\mathbf{n}(t)$ is continuous throughout the evolution. That is, $\kappa$ (and hence $\Omega$) changes sign at times $t_0$ where $\lim_{\epsilon\to0}\ddot{\mathbf{r}}(t_0+\epsilon)\cdot\ddot{\mathbf{r}}(t_0-\epsilon)<0$. The magnitude of the curvature is given by $|\kappa(t)|=\|\ddot{\mathbf{r}}(t)\|$.

\section{Robust Landau-Zener sweeps for noisy level crossings}\label{sec:zerogap}

\begin{figure}[h]
    \centering
    \includegraphics[width=1\textwidth]{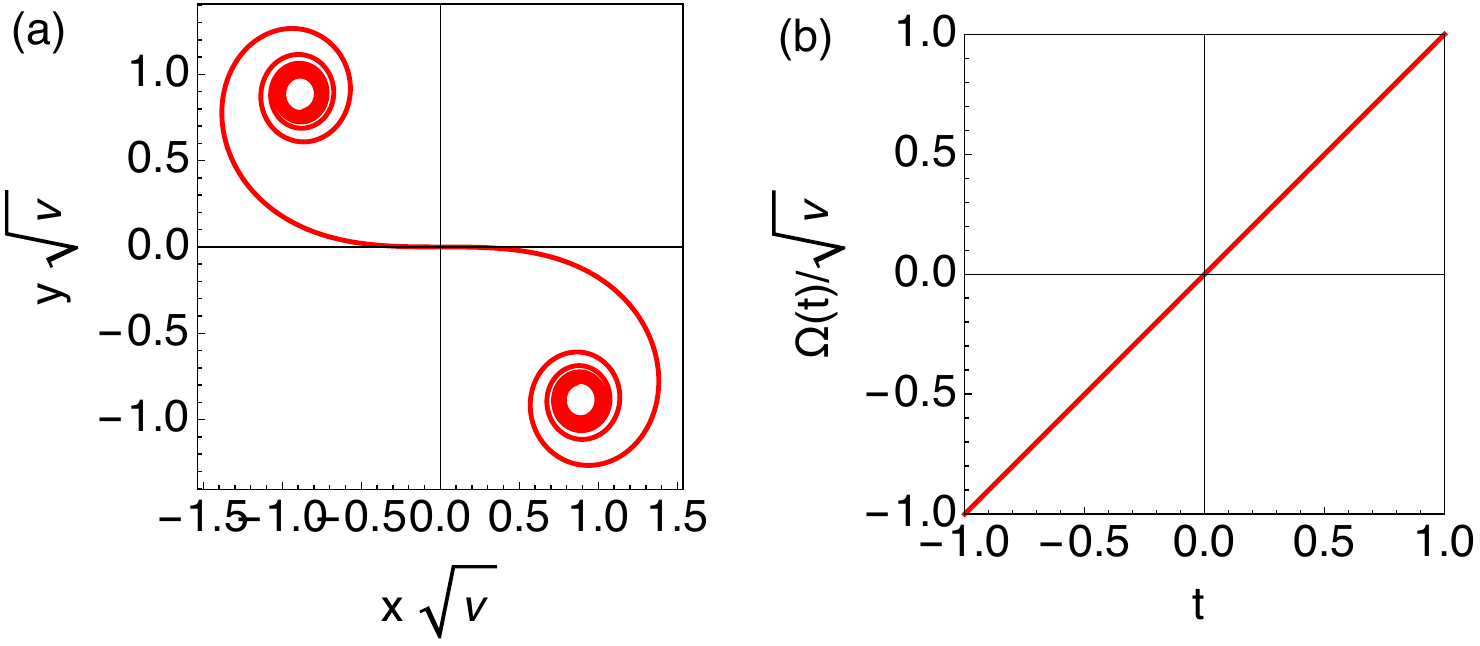}
    \caption{\textcolor{black}{The error curve (a) corresponding to a linear LZ pulse (b). In this case, the error curve is an Euler spiral.}}
    \label{fig:LZ cornu}
\end{figure}

We first consider the case where the average energy gap at the avoided crossing is zero: $\Delta=0$. In this case, the avoided crossing is due entirely to the noise fluctuation $\delta$. Since $\Delta=0$, the torsion is zero, and so the corresponding space curve $\mathbf{r}$ lives entirely in a two-dimensional plane: $\mathbf{r}(t)=[x(t),y(t),0]=[\int_0^t \cos\left(\int_0^{u} \Omega(s)ds\right)du,-\int_0^t \sin\left(\int_0^{u} \Omega(s)ds\right)du,0]$. We begin by considering the original linear LZ pulse with constant velocity, $\Omega(t)=v t$. The components of the corresponding error curve in this case are
\begin{eqnarray}\label{eq:cornu}
    x(t)&=&\int_0^t \cos(v u^2/2) du=\sqrt{\frac{2}{v}}C\left(\sqrt{\frac{v}{2}}t\right), \nonumber\\
    y(t)&=&-\int_0^t \sin(v u^2/2) du=-\sqrt{\frac{2}{v}}S\left(\sqrt{\frac{v}{2}}t\right),
\end{eqnarray}
which represents an Euler spiral curve. Here, $C(w)$ and $S(w)$ are the Fresnel integral functions. This curve is illustrated in Fig.~\ref{fig:LZ cornu}. 

\begin{figure}[h]
    \centering
    \includegraphics[width=1\textwidth]{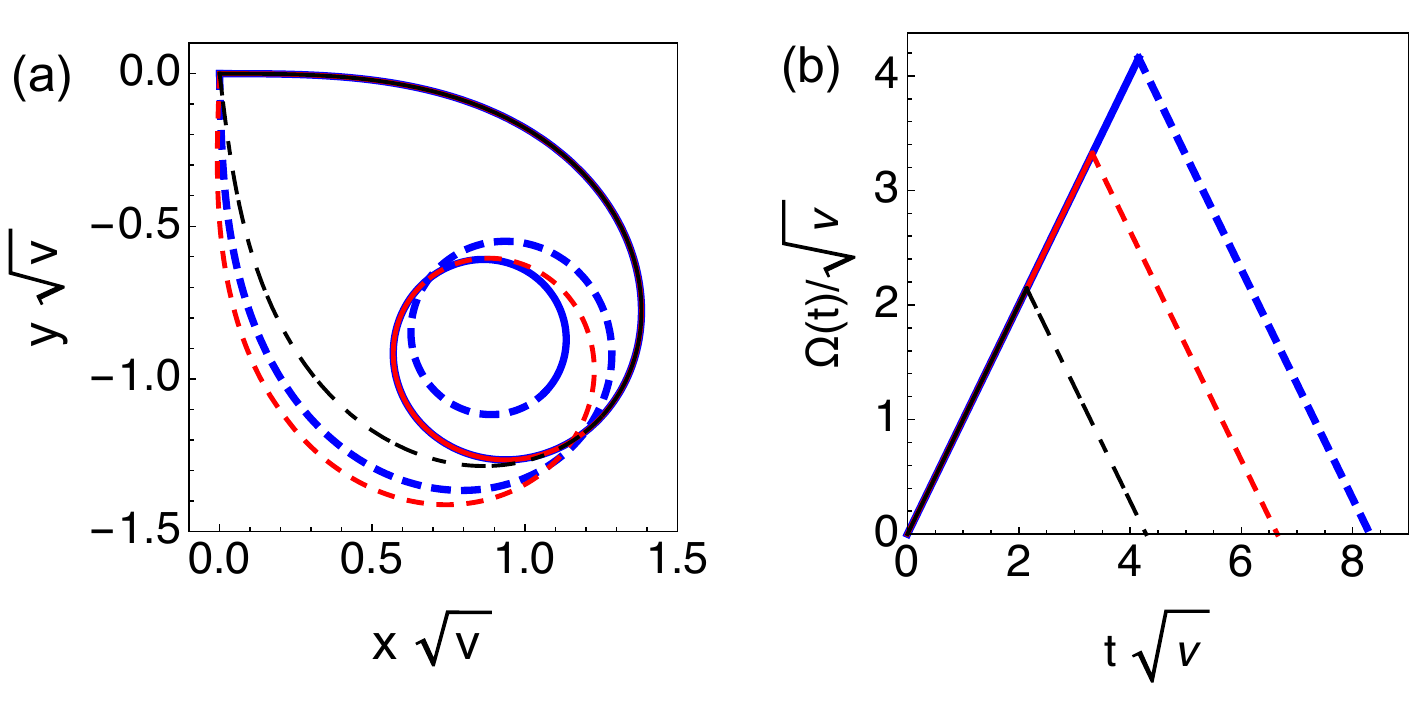}

    \caption{(a) Two Euler spiral segments can be combined to form a closed curve. The curve is traced clockwise as time evolves from $t=0$ to $t=T$. Three such curves corresponding to three different values of the sweep time $T$ are shown: \textcolor{black}{$T\sqrt{v}=2\zeta_0$ (black), $2\zeta_1$ (red), $2\zeta_2$ (blue), where $v$ is the LZ velocity, and $\zeta_0=2.14357...$, $\zeta_1=3.32193...$, $\zeta_2=4.15421...$ are the three smallest nonzero values of $t\sqrt{v}$ that solve Eq.~\eqref{eq:Cornu_closure_condition}.} (b) LZ pulses corresponding to the three curves in (a).}
    \label{fig:LZ 1st even}
\end{figure}

It is clear from Fig.~\ref{fig:LZ cornu}b that a single linear LZ sweep never corresponds to a closed error curve and thus cannot cancel noise. On the other hand, we can form a closed curve by combining two Euler spirals appropriately, as shown in Fig.~\ref{fig:LZ 1st even}a. In constructing the curve, we need to ensure that it is smooth at every point except the origin, as otherwise singularities could arise in the curvature and thus in $\Omega(t)$ \cite{zeng2018general}. If we use two Euler spiral segments to form the closed curve as in Fig.~\ref{fig:LZ 1st even}a, then smoothness is ensured by connecting the segments at a point $t$ where $\mathbf{r}(t)\cdot\dot{\mathbf{r}}(t)=0$. Using Eq.~\eqref{eq:cornu}, this condition can be rewritten as
\begin{equation}\label{eq:Cornu_closure_condition}
    \cos(vt^2/2)C(t\sqrt{v/2})+\sin(vt^2/2)S(t\sqrt{v/2})=0.
\end{equation}
There are infinitely many solutions to this equation that correspond to truncating the Euler spiral halfway through one of its infinitely many turns. The smallest nonzero solution is \textcolor{black}{$t\sqrt{v}=2.14357...\equiv\zeta_0$}, which produces the black curve shown in Fig.~\ref{fig:LZ 1st even}a after joining two such segments together. This curve yields an infinite family of LZ sweeps for which the total sweep time $T$ and the LZ velocity $v$ are related by $\sqrt{v}T=2\zeta_0$. The sweep profile $\Omega(t)$ is shown in Fig.~\ref{fig:LZ 1st even}b. The pulse starts at zero, rises linearly with velocity $|\dot\Omega|=v$ to a maximum value of $\Omega(T/2)=vT/2=\zeta_0\sqrt{v}$, and then returns to zero with the same velocity. We see that by sweeping in and out of the avoided crossing, we can cancel the first-order noise. In addition, the sweep also rotates the state about the $z$-axis by angle $vT^2/4=\zeta_0^2$.  Fig.~\ref{fig:LZ 1st even} also shows two more families of such curves (red and blue) corresponding to the next two smallest solutions, \textcolor{black}{$\zeta_1=3.32193...$ and $\zeta_2=4.15421...$}, of Eq.~\eqref{eq:Cornu_closure_condition}. These curves contain additional loops, while the associated pulses are longer and of higher amplitude. These pulses generate $z$ rotations by angles $\zeta_1^2$ and $\zeta_2^2$, respectively.

\begin{figure}[h]
    \centering
    \includegraphics[width=\textwidth]{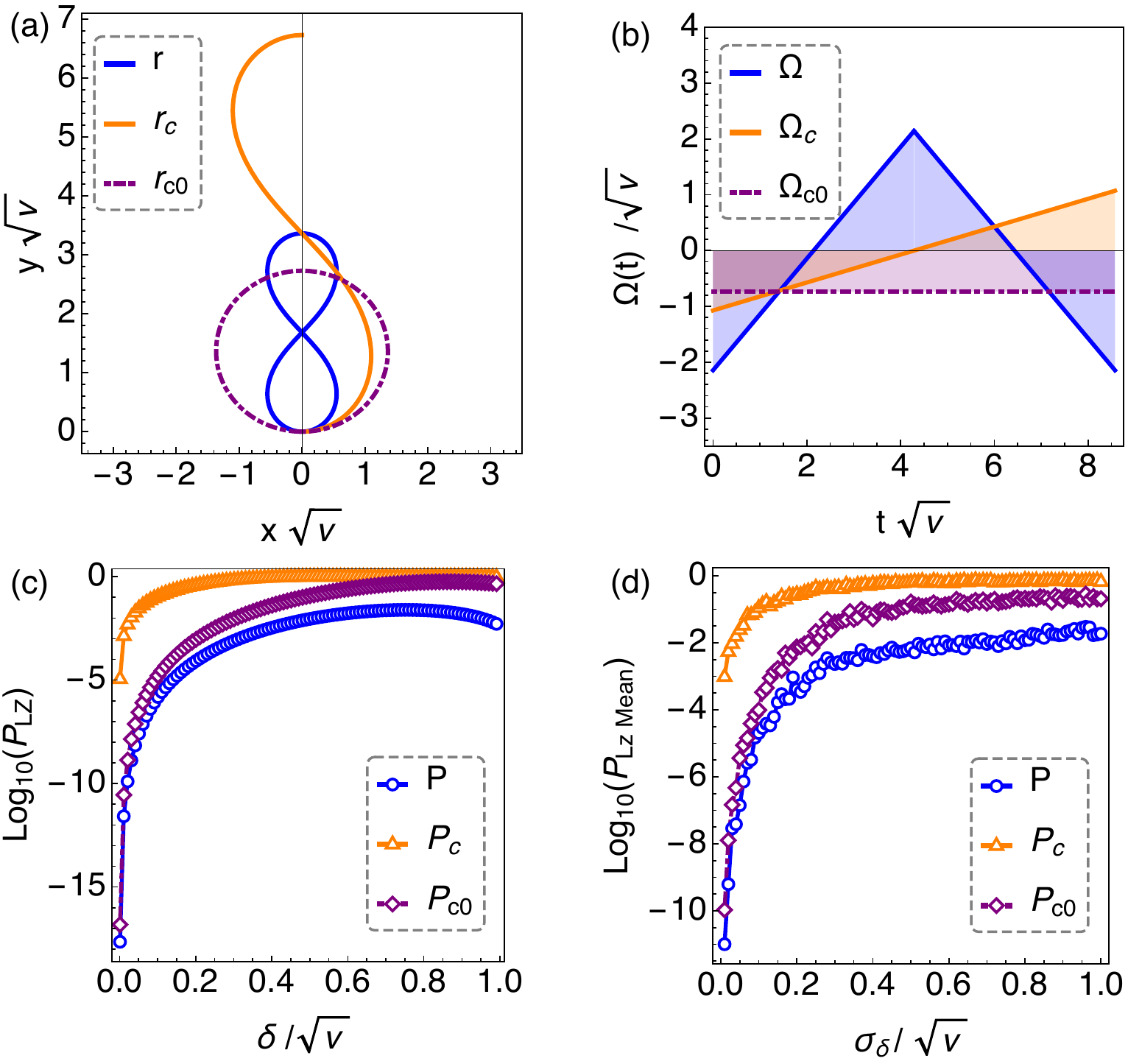}
    \caption{(a) Closed curve with zero enclosed area formed by combining two Euler spiral segments (blue), a closed curve with nonzero area (purple), and an open segment of Euler spiral (orange). All correspond to identity operations. The total arc length of each curve is \textcolor{black}{$T=4\zeta_0/\sqrt{v}$, where $v$ is the LZ velocity, and $\zeta_0=2.14357...$ is the smallest nonzero value of $t\sqrt{v}$ that solves Eq.~\eqref{eq:Cornu_closure_condition}.}
    (b) The LZ sweep profiles corresponding to the three curves in (a). 
    (c) The LZ transition probability as a function of the noise error $\delta$ for the three pulses shown in (b). (d) The average LZ transition probability as a function of the noise strength $\sigma_\delta$. For each value of $\sigma_\delta$, $P_{\mathrm{LZ}}$ is averaged over 100 instances of $\delta$ randomly sampled from a normal distribution with zero mean and standard deviation $\sigma_\delta$.}
    \label{fig:evenSweepAllPlots}
\end{figure}

This procedure can be generalized to canceling second-order noise errors. This requires making the error curve not only closed but also such that it encloses zero area. \textcolor{black}{We can construct a zero-area curve by again piecing together two Euler spiral segments}, but this time using longer segments to form a figure-8 shape as shown in Fig.~\ref{fig:evenSweepAllPlots}a (blue curve). This curve can be thought of as two copies of the curve from Fig.~\ref{fig:LZ 1st even} glued together, except here the curve starts and ends at the bottom of the figure-8 rather than in the center. Because the curve starts and ends at a point of nonzero curvature, this corresponds to the system starting and ending away from the level crossing. The sweep profile $\Omega(t)$ is shown in Fig.~\ref{fig:evenSweepAllPlots}b, where it is clear that the system is tuned through the crossing twice before returning to its initial configuration. This sweep implements an identity operation as can be seen from the fact that the pulse area is zero. \textcolor{black}{It is also possible to construct curves corresponding to other types of gates that are robust to second order. An example of a different family of zero-area curves formed by joining straight lines and circular arcs is given in Appendix~\ref{Appendix:phase gate}. This family can generate any single-qubit phase gate. Because the curve is comprised of straight lines and circular arcs, the corresponding pulses are square pulse sequences in this case. Such square pulse sequences can be transformed into smooth continuous pulses by smoothing the error curves. A smoothing method is described in detail below.}

To test the noise-robustness of the LZ pulse shown in Fig.~\ref{fig:evenSweepAllPlots}b, we numerically compute the transition probability $P_{\mathrm{LZ}}$ as a function of the noise parameter $\delta$. Here, $P_{\mathrm{LZ}}$ is defined as the probability to transition from one diabatic state to the other: $P_{\mathrm{LZ}}=|\langle 1|U(T)|0\rangle|^2$. The results are shown in Fig.~\ref{fig:evenSweepAllPlots}c. For comparison, we also show results for a linear LZ sweep, $\Omega_c(t)$, that does not correct for noise errors and a constant pulse, $\Omega_{c0}$, that corrects noise to first order. These pulses and their associated curves are shown in panels (a) and (b) (orange and purple curves). It is clear from panel (a) that the curve associated with the linear sweep does not close, and hence the first-order noise error is not eliminated. On the other hand, the curve for the constant pulse does close for the chosen pulse area. It is clear from Fig.~\ref{fig:evenSweepAllPlots}c that the geometrically designed $\Omega(t)$ reduces $P_{\mathrm{LZ}}$ by orders of magnitude compared to the linear sweep; it outperforms the constant pulse as well, especially for larger noise strengths. In reality, the probability $P_{\mathrm{LZ}}$ is obtained by averaging over many runs of the experiment, and the value of $\delta$ can vary from one run to the next. To account for this, we average $P_{\mathrm{LZ}}$ over 100 values of $\delta$ randomly chosen from a normal distribution with zero mean and standard deviation $\sigma_\delta$. The resulting noise-averaged $P_{\mathrm{LZ}}$ as a function of $\sigma_\delta$ is shown in Fig.~\ref{fig:evenSweepAllPlots}d for the geometrically engineered pulse, the linear sweep, and the constant pulse. We see that the engineered pulse $\Omega(t)$ yields a $P_{\mathrm{LZ}}$ that is 2-5 orders of magnitude smaller than that of $\Omega_c(t)$ and up to 1.5 orders of magnitude smaller than that of $\Omega_{c0}$ depending on the noise strength $\sigma_\delta$.

\begin{figure}[H]
    \centering
    \includegraphics[width=0.98\textwidth]{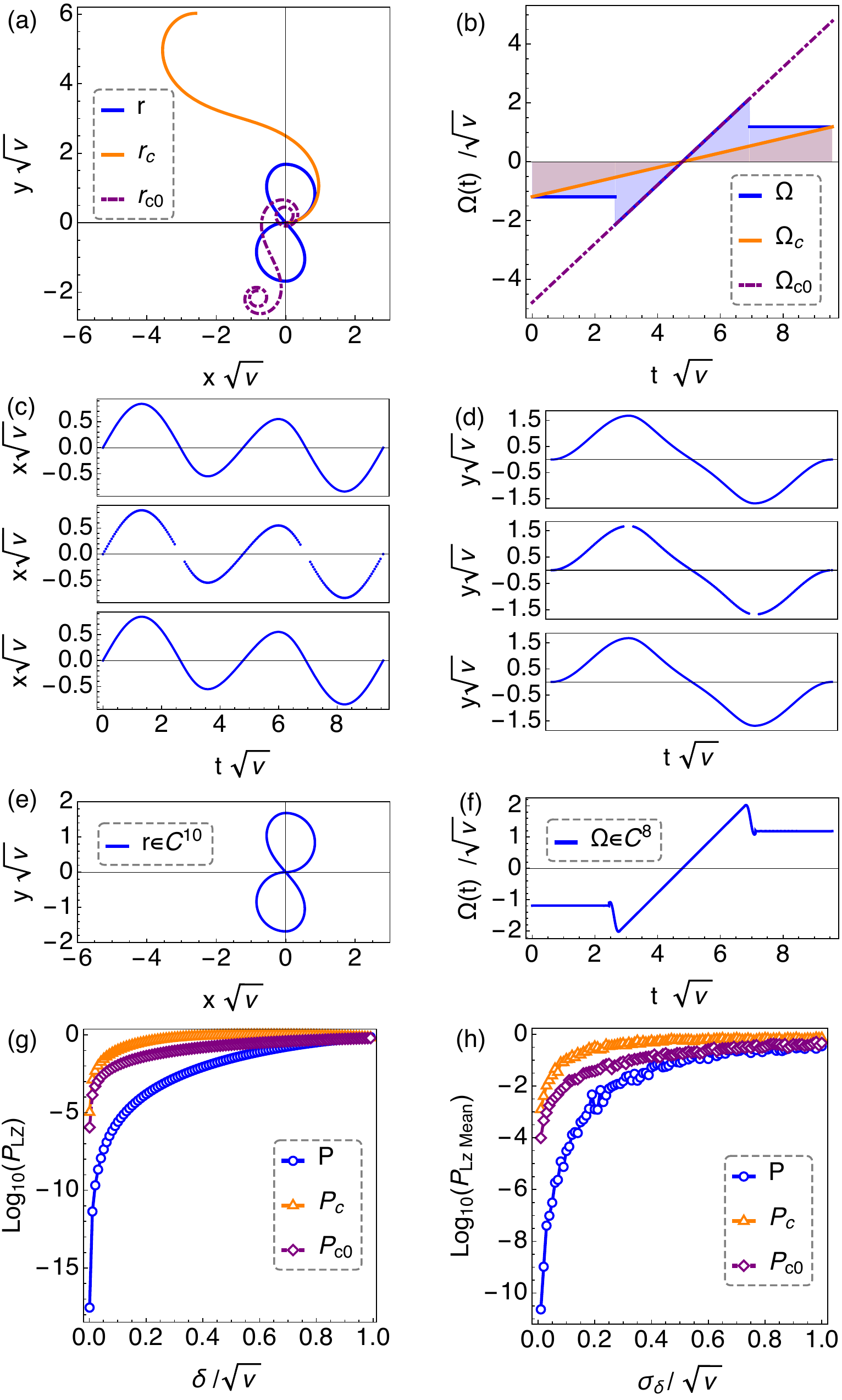}
   \caption{\textcolor{black}{(a) Closed curve with zero net area formed by joining a segment of Euler spiral with two semicircles (blue). Two curves corresponding to uncorrected linear sweeps are also shown (orange and purple). (b) LZ pulses corresponding to the curves in (a). (c) From top to bottom: original function $x(t)\in C^1$; discretized version with step size $dt=0.01/\sqrt{v}$ and points removed; fitted $C^{10}$ curve. (d) Same curve fitting procedure for $y(t)$ as in (c). (e) Smoothed $C^{10}$ error curve. (f) Continuous and smooth ($C^{8}$) noise-robust pulse derived from error curve in (e). (g) The LZ transition probability as a function of the noise error $\delta$ for the naive sweeps (orange and purple) and the noise-cancelling sweep (blue). (d) The average LZ transition probability for the noise-cancelling sweep (blue) and naive sweeps (orange and purple) versus noise strength $\sigma_\delta$. For each value of $\sigma_\delta$, $P_{\mathrm{LZ}}$ is averaged over 100 instances of $\delta$ randomly sampled from a normal distribution with zero mean and standard deviation $\sigma_\delta$.}}
   \label{fig:oddSweepLZAllPlots}
\end{figure}

What if we want to tune the system through the noisy crossing once such that $\Omega(0)<0$ and $\Omega(T)>0$? In this case, the curve must start with a negative value of curvature and end with a positive value. This can be accomplished by combining a segment of Euler spiral with two semicircles, as shown in Fig.~\ref{fig:oddSweepLZAllPlots}a (blue curve). Here, one semicircle is traced first as time evolves from $t=0$ to $t=T_{\mathrm{circ}}$, where $T_{\mathrm{circ}}$ is the length of the semicircle.
\noindent Then the half-figure-8 segment of Euler spiral is traced from $t=T_{\mathrm{circ}}$ to $t=T_{\mathrm{circ}}+T_{\mathrm{Euler}}$, where we already know from the example in Fig.~\ref{fig:evenSweepAllPlots} that $T_{\mathrm{Euler}}=2\zeta_0/\sqrt{v}$. Finally, the second semicircle is traced from $t=T_{\mathrm{circ}}+T_{\mathrm{Euler}}$ to $t=2T_{\mathrm{circ}}+T_{\mathrm{Euler}}=T$. In order to form a smooth closed curve, we need the diameter, $d$, of the semicircles to equal the maximal distance from the origin reached by the Euler spiral:
\begin{equation}
    d=\sqrt{(2/v)[C(\zeta_0/\sqrt{2})]^2+(2/v)[S(\zeta_0/\sqrt{2})]^2}\approx1.68/\sqrt{v}.
\end{equation}
Each semicircle has a constant curvature given by $\Omega_{\mathrm{circ}}=2/d\approx1.19\sqrt{v}$ and length $T_{\mathrm{circ}}=\pi d/2\approx2.64/\sqrt{v}$. These segments of the curve thus correspond to holding the drive $\Omega$ constant for a time $T_{\mathrm{circ}}$ at the beginning and end of the LZ sweep. The segment of Euler spiral contributes an increasing linear ramp to the drive that starts from $\Omega(T_{\mathrm{circ}})=-\sqrt{v}\zeta_0$ (the maximum curvature of the Euler spiral segment) and ends at $\Omega(T_{\mathrm{circ}}+T_{\mathrm{Euler}})=\sqrt{v}\zeta_0$, with constant LZ velocity $|\dot\Omega|=v$. The complete sweep profile, $\Omega(t)$, is shown in Fig.~\ref{fig:oddSweepLZAllPlots}b.

\textcolor{black}{The error-correcting LZ pulse shown in Fig.~\ref{fig:oddSweepLZAllPlots}b exhibits two jump discontinuities, which can make it challenging to implement in experiments. These discontinuities are a direct result of the fact that the error curve in Fig.~\ref{fig:oddSweepLZAllPlots}a is $C^1$, i.e., its first derivative is continuous, but not its second derivative. However, this error curve is $C^{\infty}$ (infinitely differentiable) at all but two points, namely the two joining points at which the semicircles meet the Euler spiral. These two $C^1$ points can be eliminated by applying a smoothing procedure in which the whole curve is fit to a $C^M$ curve, where we need $M=2$ to obtain a continuous pulse, and $M\geq3$ guarantees the pulse is not only continuous but also smooth to order $C^{M-2}$. Such a smoothing procedure can be implemented in many different ways. Here, we discretize the curve into a set of points separated in arc length by step size $dt$, we delete a few points in the vicinity of the $C^1$ points, and we then fit the remaining points to a polynomial of degree $M$. The resulting fitted curve is in differentiability class $C^M$. The application of this procedure to the error curve in Fig.~\ref{fig:oddSweepLZAllPlots}a is depicted in Figs.~\ref{fig:oddSweepLZAllPlots}c,d, which show the original curve coordinates $x(t)$ and $y(t)$, their discretization, and the subsequent fits. The fitting is done using the built-in command ``Interpolation" in Mathematica. The interpolation order $M$ and the number of eliminated points $N_{el}$ are adjustable parameters. In this example, we choose $M=10$, $N_{el}=7$ to smooth the pulse without losing too much of its original shape. The resulting curve and corresponding pulse are shown in Figs.~\ref{fig:oddSweepLZAllPlots}e,f.} Since the curve we constructed is not only closed but also has zero net enclosed area, this LZ pulse cancels noise to second order. This is confirmed in Figs.~\ref{fig:oddSweepLZAllPlots}g,h, where we again observe a reduction in the transition probability by several orders of magnitude compared to naive linear sweeps.

A notable feature of the robust LZ sweep we just constructed is that it varies non-monotonically from $\Omega(0)=-1.19\sqrt{v}$ to $\Omega(T)=1.19\sqrt{v}$ (see Fig.~\ref{fig:oddSweepLZAllPlots}b). This non-monotonicity is in fact inevitable. As we prove in Appendix~\ref{app:Kneser}, it is impossible to construct a monotonic sweep that cancels noise, even to first order. This is a direct consequence of a result from differential geometry known as the Tait-Kneser nesting theorem. This is yet another example of how well-known constructs from differential geometry can shed light on the physics of LZ transitions. \textcolor{black}{In Appendix~\ref{app:scaling}, we further show that the LZ probability associated with the geometrically engineered pulse in Fig.~\ref{fig:oddSweepLZAllPlots}f exhibits a qualitatively distinct dependence on $\delta^2/v$ compared to the exponential decay exhibited by the original Landau-Zener formula for an infinite, linear sweep.}

\section{Robust Landau-Zener sweeps for noisy avoided crossings}\label{sec:nonzerogap}

We now consider the case where the avoided crossing has a nonzero energy gap even in the absence of noise: $\Delta>0$. As discussed above, the evolution is now described by space curves in three dimensions with constant torsion given by $\tau=-\Delta$. To cancel first-order noise, the curve must also be closed. Constructing such curves is significantly more challenging than finding closed plane curves. Here, we present a systematic approach for obtaining such curves.

We begin by writing the curve in terms of the binormal vector only. This can be done by first writing the tangent vector in terms of the normal and binormal vectors: $\dot{\mathbf{r}}=\mathbf{n}\times\mathbf{b}$. We then use the Frenet-Serret equations (Eq.~\eqref{eq:FSframe}) to replace $\mathbf{n}\rightarrow-1/\tau\dot{\mathbf{b}}$. We thus obtain
\begin{equation}\label{eq:rFromb}
    \mathbf{r}(t)=\frac{1}{\tau}\int_0^t\mathbf{b}(s)\times\dot{\mathbf{b}}(s)ds.
\end{equation}
\textcolor{black}{Here, we used the fact that the torsion is constant to bring it outside the integral. Notice that this integral has the same mathematical form as the second-order error cancellation constraint in Eq.~\eqref{A2 parametric}. An important difference between Eq.~\eqref{eq:rFromb} and Eq.~\eqref{A2 parametric} is that $\mathbf{b}(s)$ is a spherical curve, as the binormal vector is defined to be normalized. Eq.~\eqref{eq:rFromb} tells us that we can construct a closed curve of constant torsion by finding a spherical curve with zero area.}

\textcolor{black}{Recall that we must also satisfy the condition $\|\dot{\mathbf{r}}(t)\|=1$; this requires the tangent vector of $\mathbf{b}$ to be normalized to $|\tau|$:
\begin{eqnarray}\label{eq:btangent}
    \|\dot{\mathbf{r}}(t)\|=\frac{1}{|\tau|}\|\mathbf{b}(t)\times\dot{\mathbf{b}}(t)\|=\frac{1}{|\tau|}\|\dot{\mathbf{b}}(t)\|=1.
\end{eqnarray}
\begin{figure}[H]
        \centering
        \includegraphics[width=0.9\textwidth]{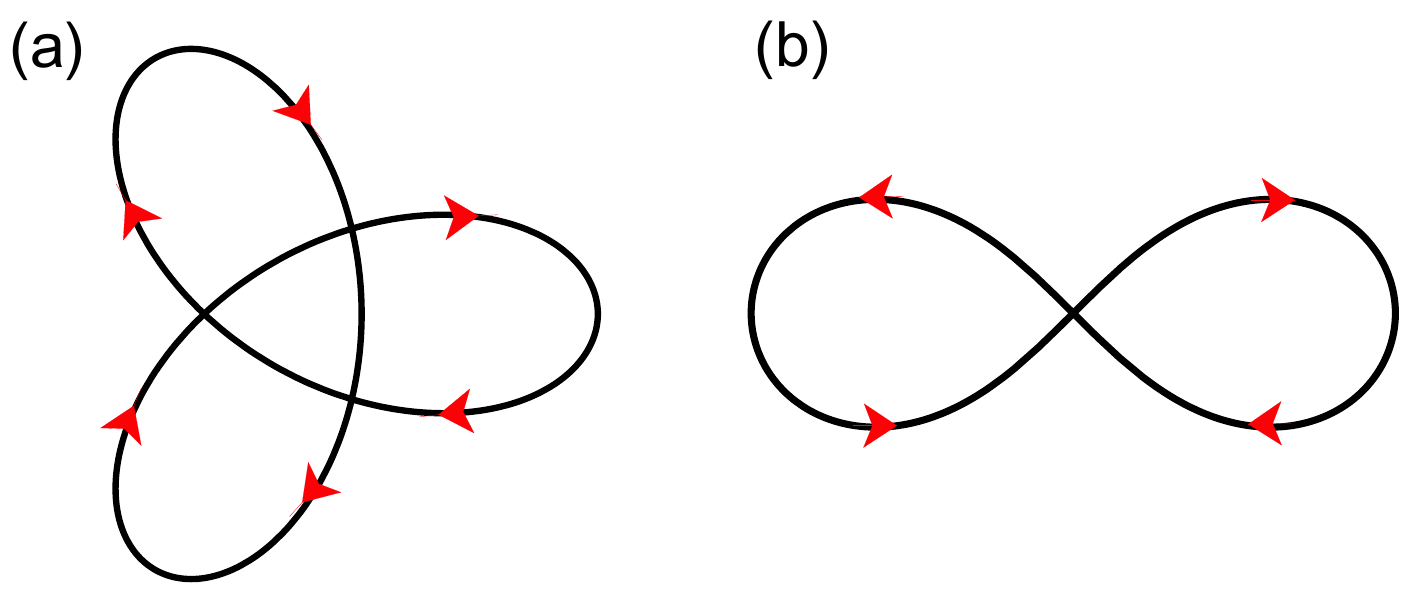}
        \caption{(a) Example of a critically three-fold-rotationally symmetric plane curve. (b) A rotationally symmetric curve that is not critically rotationally symmetric. All the arrows are reversed under an in-plane $\pi$ rotation about the center.}
        \label{fig: critically symmetric}
    \end{figure}
\noindent Instead of building this condition directly into $\mathbf{b}(s)$, we will instead design a binormal curve with non-constant velocity, compute $\widetilde{\mathbf{r}}(s)$ from Eq.~\eqref{eq:rFromb}, and then obtain $s(t)$ by solving $t=\int_0^s \|\dot{\widetilde{\mathbf{r}}}(s')\| d s'$. The error curve with properly normalized velocity is then given by $\mathbf{r}(t)=\widetilde{\mathbf{r}}(s(t))$. An advantage of this approach is that the same $\mathbf{b}(s)$ can be used to construct a constant-torsion curve for any value of the torsion $\tau$. This is true because $\tau$ does not enter into the unit-sphere or vanishing-area conditions that $\mathbf{b}(s)$ must satisfy; it only enters in after $\mathbf{b}(s)$ is constructed, through Eq.~\eqref{eq:rFromb}.}

To design a curve on the unit sphere with vanishing-area projections, we employ the following general strategy \cite{weiner1977closed}. To begin with, a systematic way to construct a curve on a unit sphere is to first choose a curve in the $xy$ plane, $b^{\hat{z}}(s)=[x(s),y(s)]$, and then project this onto the northern hemisphere: $\mathbf{b}(s)=[x(s),y(s),\sqrt{1-x^2(s)-y^2(s)}]$. This works provided $b^{\hat{z}}(s)$ lies within a unit disk. The question then becomes: How can we construct $b^{\hat{z}}(s)$ such that $\mathbf{b}(s)$ satisfies the vanishing-area conditions? The area of the projection onto the $xy$ plane can be cancelled by choosing $b^{\hat{z}}(s)$ such that it encloses zero area. The remaining two projected areas can be made to vanish by choosing $b^{\hat{z}}(s)$ to possess a discrete rotational symmetry about the $z$-axis. $\mathbf{b}(s)$ will inherit this symmetry and, if the symmetry is chosen appropriately, the areas of the projections onto the $xz$ and $yz$ planes will vanish. Importantly, the plane curve must be {\it critically} rotationally symmetric in order for this approach to work, meaning that after a symmetry operation is performed, not only must the curve be preserved, but also the manner in which it is traversed. This distinction is illustrated in Fig.~\ref{fig: critically symmetric}.

An example of a closed curve of constant torsion with a three-fold rotational symmetry was obtained using these methods in Ref.~\cite{weiner1977closed}. This space curve was subsequently employed to design a control pulse that cancels second-order noise in Ref.~\cite{zeng2019geometric}. This example yields a pulse that starts and ends at zero. Here, we are instead interested in a LZ sweep from one side of the avoided 

\begin{figure}[H]
\centering
     \includegraphics[width=1.9\columnwidth]{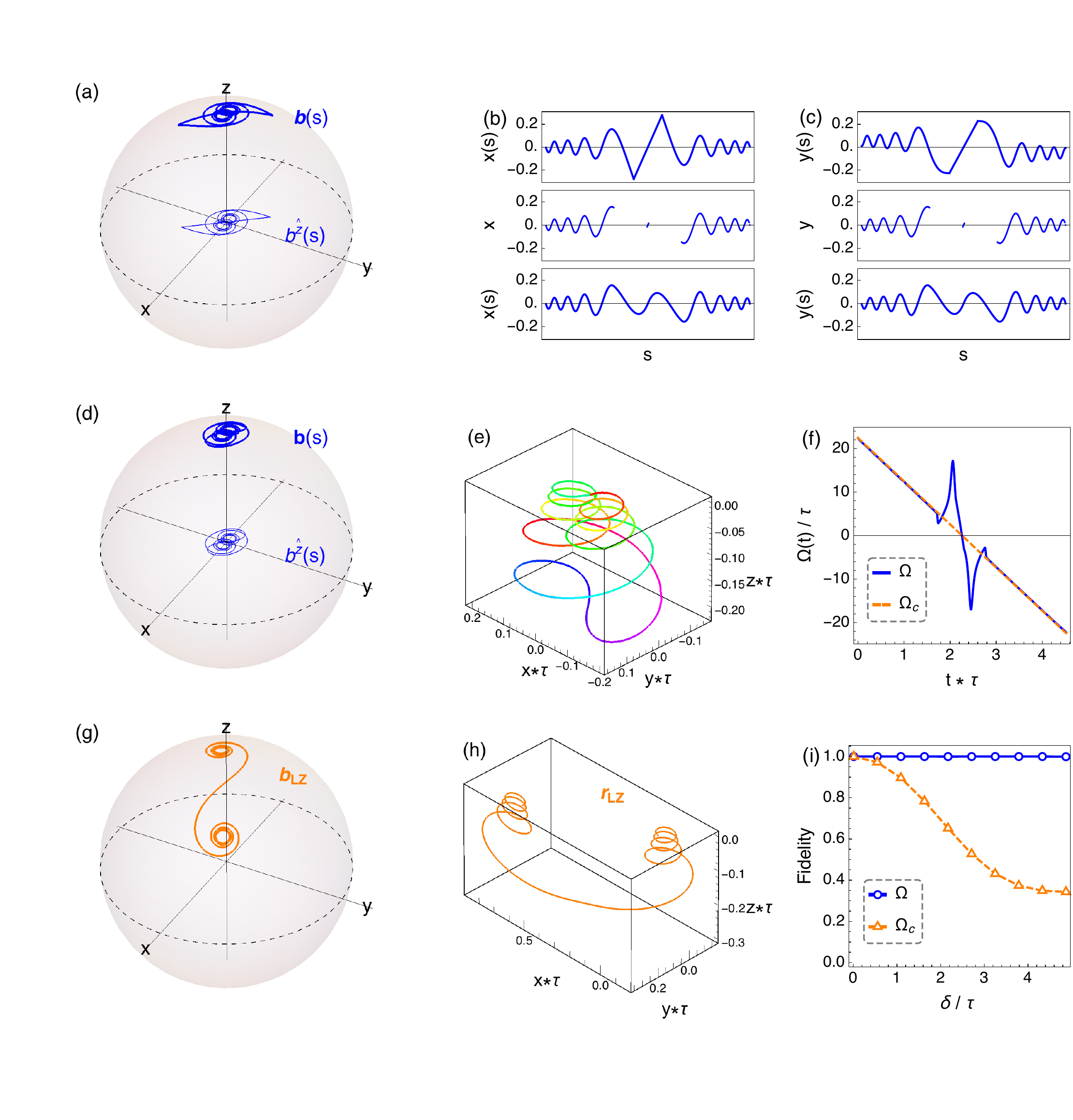}
\captionsetup{width=\columnwidth}
    \caption{\textcolor{black}{(a) Plane curve $b^{\hat{z}}(s)$ and its projection $\mathbf{b}(s)$ onto the unit sphere, which is the binormal curve. Both curves satisfy zero-area conditions. (b) From top to bottom: original $x(s)\in C^0$; discretized version of $x(s)$ with points near cusps removed; $C^3$ fit. (c) Same as in (b) but for $y(s)$. (d) $C^3$ binormal curve produced from smoothing procedure and with zero area in all three Cartesian directions. (e) Closed constant-torsion ($\tau$) space curve obtained from smoothed binormal curve in (d) using Eq.~\eqref{eq:rFromb}. (f) The resulting robust LZ pulse (blue) obtained from the curvature of the curve in (e), and a naive linear pulse (orange) included for comparison. (g) Binormal curve and (h) constant-torsion space curve for naive linear LZ pulse in (f). (i) The fidelity of the operations implemented by the pulses in (f) relative to the ideal, noiseless case.}
     }\label{fig:nonzeroallplots}
\end{figure}

~
\newpage

\noindent crossing to the other such that $\Omega(0)>0$ and $\Omega(T)<0$. This requires a curve with less symmetry.

\textcolor{black}{We solve this problem by starting with a plane curve with two-fold rotational symmetry, comprised of two Euler spiral segments connected by a straight line passing through the origin. Each Euler spiral segment is parameterized as 
\begin{align}
    b_{\text{Euler}}^{\hat{z}}(s)&=\left[\frac{1}{\sqrt{5}}C\left(\sqrt{\frac{v}{2}}(t_f-s)\right)+x_a\right]\hat x\nonumber+\\
    &\left[\frac{1}{\sqrt{5}}S\left(\sqrt{\frac{v}{2}}(t_f-s)\right)+y_a\right]\hat y, 
\end{align}
where $v $ is the LZ velocity, and $t_f$ is the length of the Euler spiral segment. The resulting $b^{\hat{z}}(s)$, along with its projection $\mathbf{b}(s)$ onto the unit sphere, is shown in Fig.~\ref{fig:nonzeroallplots}a, where it is evident that both curves exhibit a two-fold rotational symmetry. However, they are not critically rotationally symmetric, which means that the strategy described above cannot be fully applied here. Another thing to notice from Fig.~\ref{fig:nonzeroallplots}a is that $\mathbf{b}(s)$ has two sharp corners. These two cusps will result in two singularities in the curvature and hence two delta pulses in $\Omega(t)$. To eliminate these cusps, we apply the same smoothing procedure we used above for Fig.~\ref{fig:oddSweepLZAllPlots}; the difference here is that we smooth the binormal curve instead of the error curve directly. For the spherical curve \(\mathbf{b}(s)=[x(s),y(s),\sqrt{1-x(s)^2-y(s)^2}]\), we just need to smooth the functions $x(s)$ and $y(s)$, and the third component $\sqrt{1-x(s)^2-y(s)^2}$ will automatically be smoothed as long as $b^{\hat{z}}(s)$ lies within the unit disk. This smoothing process is somewhat more challenging than that used in Fig.~\ref{fig:oddSweepLZAllPlots}, because the smoothing procedure alters the shape of the binormal curve, and thus changes the area it encloses. This means that a new set of parameters $\{t_f,x_a,y_a\}$ must be picked in each iteration of the smoothing process to ensure the vanishing-area condition is maintained. In this example, we choose the discretization step size to be $d s=0.01/\tau$, interpolation order $M=3$ and, because we need to smooth sharp cusps, the number of eliminated dots is $N_{el}=85$, with 50 points on the relatively steeper Euler spiral side and 34 on the great arc side of each cusp (the great arc being the projection of the straight line segment onto the unit sphere). For $v=10\tau^2$, we numerically calculated $\{t_f,x_a,y_a\}=\{2.2237391/\tau,-0.2800000,-0.2280100\}$ such that the smoothed curve $\mathbf{b}(s)$ satisfies the condition $\int\mathbf{b}\times d\mathbf{b}=\mathbf{0}$ to a precision of $10^{-7}$.}

\textcolor{black}{With a suitable $\mathbf{b}(s)$ at hand, we next perform the integration in Eq.~\eqref{eq:rFromb} followed by the arc length reparameterization procedure to obtain the desired constant-torsion space curve $\mathbf{r}(t)$, which is shown in Fig.~\ref{fig:nonzeroallplots}e. Computing the curvature of this curve yields the LZ pulse profile $\Omega(t)$, which is shown in Fig.~\ref{fig:nonzeroallplots}f. The initial and final pieces of the pulse are approximately linear ramps that stem from the Euler spiral segments (the ramps are not exactly linear because $\mathbf{b}_{\mathrm{Euler}}$ is not the real binormal curve of a curve with linear curvature),  while the nontrivial behavior in between is a reflection of the fitted version of the great arc. We again see the robust pulse is non-monotonic, which is consistent with the theorem in Appendix~\ref{app:Kneser} for plane curves. 
To confirm that the LZ sweep we constructed is indeed noise-resistant, we show the fidelity as a function of $\delta$ in Fig.~\ref{fig:nonzeroallplots}i. For comparison, we also show the fidelity for a perfectly linear sweep $\Omega_c(t)$ that has the same LZ velocity as the linear segments of our geometrically engineered pulse. The duration of $\Omega(t)$ and $\Omega_c(t)$ is $t_{final}=4.51727*\tau$, with \(\Omega_c(t)=-9.9175 (t-\frac{t_{final}}{2})*\tau^2\). The binormal and error curves that produce this LZ sweep are shown in Figs.~\ref{fig:nonzeroallplots}g,h. A comparison of  fidelities in Fig.~\ref{fig:nonzeroallplots}i reveals that the geometrically engineered sweep provides a substantial improvement compared to the uncorrected sweep over a broad range of noise fluctuation strengths.
}

\section{Conclusions }\label{sec:conclusion}
In conclusion, we presented a general approach to constructing noise-resistant Landau-Zener sweeps. This approach exploits and extends a recently discovered connection between qubit evolution and space curves in three dimensions. In the case where the avoided crossing is due purely to noise fluctuations, we showed how to cancel noise errors up to second order by drawing closed plane curves of zero net area. This can be done regardless of whether the sweep starts and ends on the same or opposite sides of the avoided crossing. We further proved that it is impossible to suppress noise with a monotonic sweep. In the case where the avoided crossing has a finite gap in the absence of noise, we showed that robust Landau-Zener sweeps correspond to closed curves of constant torsion. We presented a general method to construct such curves and demonstrated it by deriving explicit examples of noise-resistant sweeps. In all cases, we confirmed the cancellation of noise via numerical simulations. Our results constitute a general recipe for exploiting avoided crossings despite the presence of noise.

\section*{Acknowledgments}
This work was supported by the Department of Energy (grant no. DE-SC0018326). E.B. also acknowledges support by the Office of Naval Research (grant no. N00014-17-1-2971).


\appendix

\section{Arbitrary phase gates robust to second order}\label{Appendix:phase gate}

\begin{figure}
    \centering
    \includegraphics[width=1\textwidth]{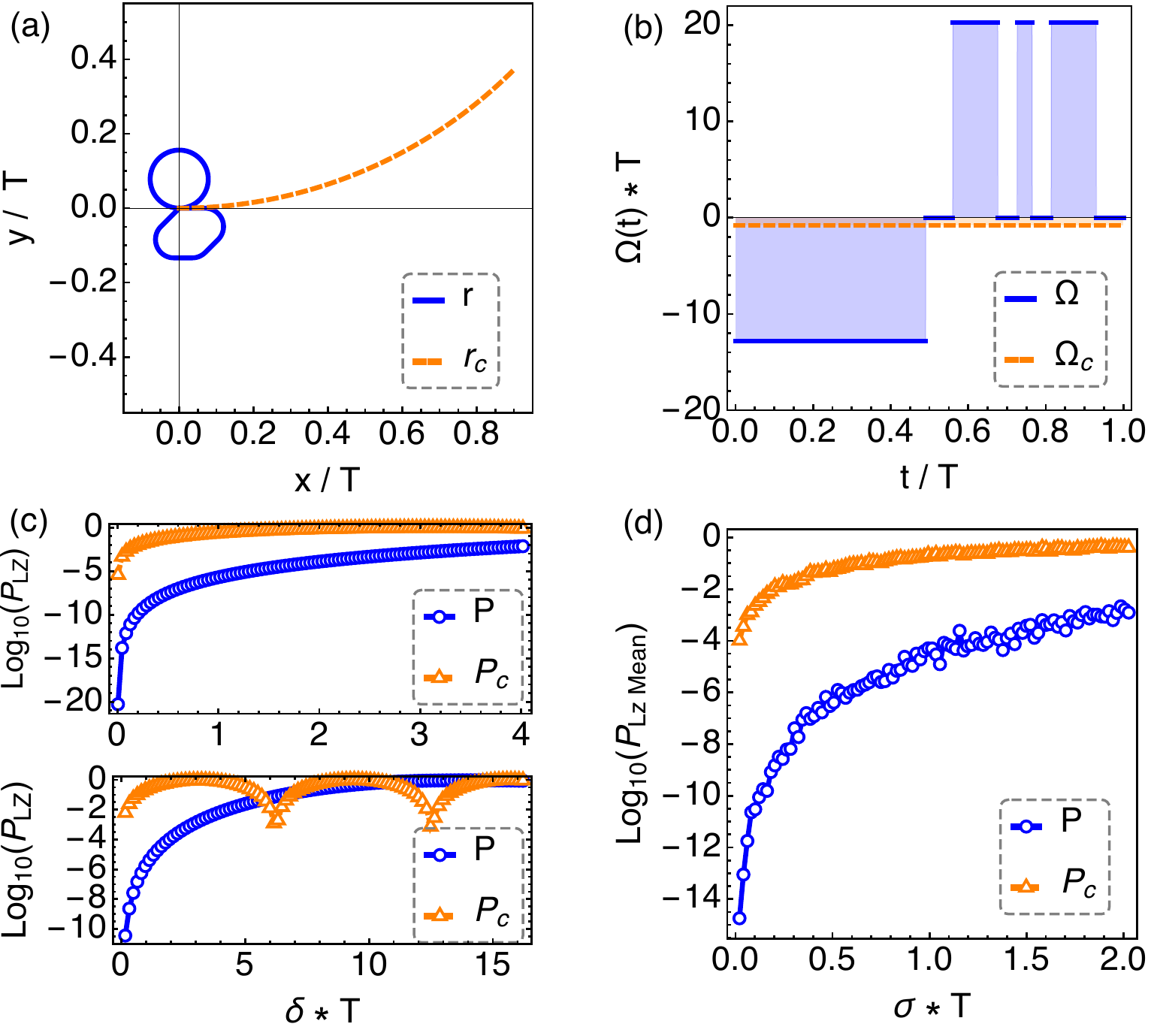}
    \caption{Designing a $\pi/4$ phase gate robust to second order. (a) Closed plane curve with vanishing enclosed area (blue) and curve corresponding to uncorrected square pulse (orange). (b) The pulses corresponding to the curves in (a). (c) The LZ transition probability as a function of the noise error $\delta$ for the pulses in (b). The upper panel is a zoom-in of the lower one. (d) The average LZ transition probability for the pulses in (b) as a function of the noise strength $\sigma_\delta$. For each value of $\sigma_\delta$, $P_{\mathrm{LZ}}$ is averaged over 100 instances of $\delta$ randomly sampled from a normal distribution with zero mean and standard deviation $\sigma_\delta$.}
   \label{fig:phaseplots}
\end{figure}

\begin{figure}
    \centering
    \includegraphics[width=1\textwidth]{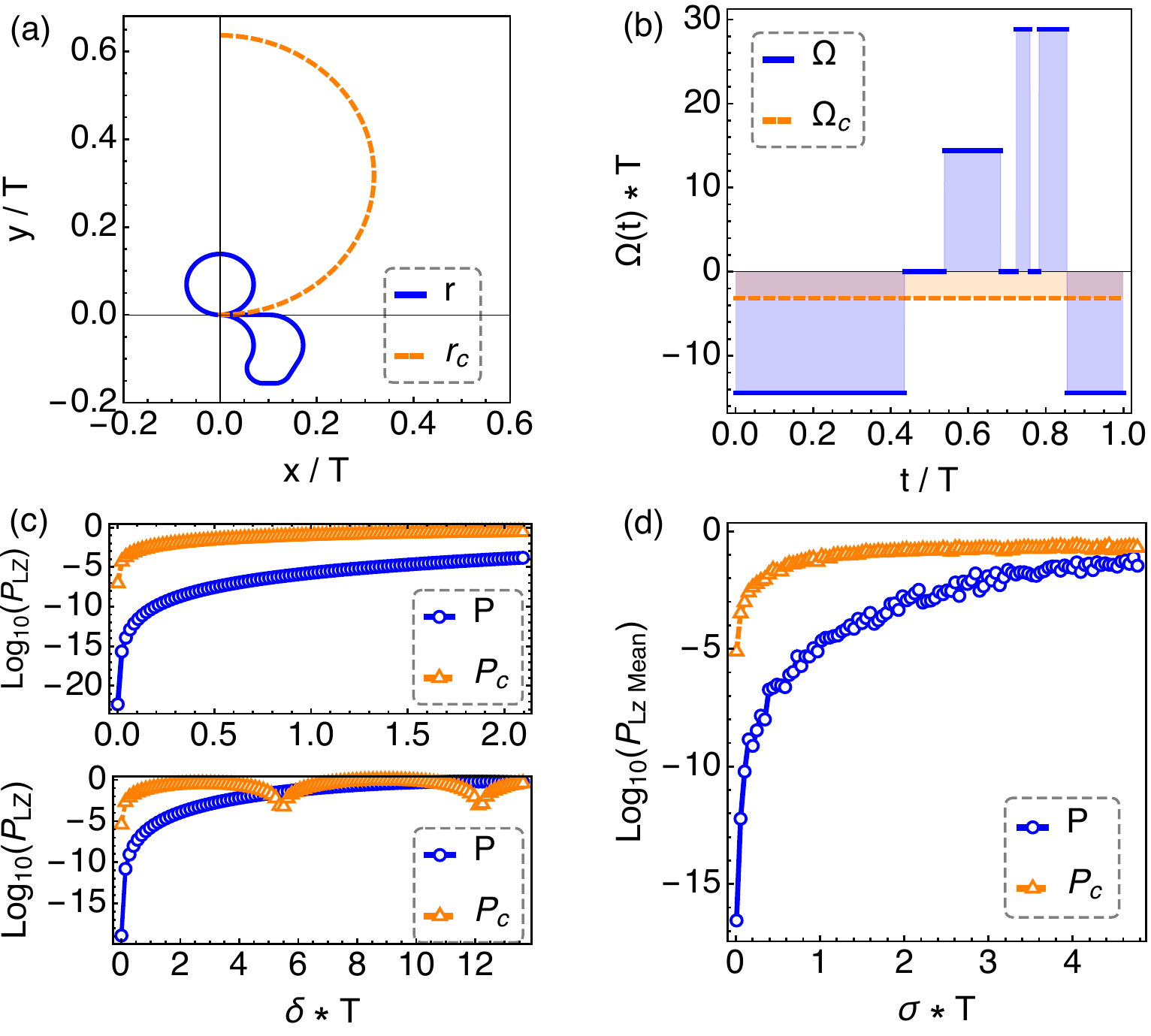}
    \caption{Designing a $\pi$ phase gate robust to second order. (a) Closed plane curve with vanishing enclosed area (blue) and curve corresponding to uncorrected square pulse (orange). (b) The pulses corresponding to the curves in (a). (c) The LZ transition probability as a function of the noise error $\delta$ for the pulses in (b). The upper panel is a zoom-in of the lower one. (d) The average LZ transition probability for the pulses in (b) as a function of the noise strength $\sigma_\delta$. For each value of $\sigma_\delta$, $P_{\mathrm{LZ}}$ is averaged over 100 instances of $\delta$ randomly sampled from a normal distribution with zero mean and standard deviation $\sigma_\delta$.}
   \label{fig:phaseplots2}
\end{figure}

Here, we provide an example of a family of zero-area plane curves that yield second-order robust phase gates. Representatives of this family are shown in Figs.~\ref{fig:phaseplots}a and \ref{fig:phaseplots2}a (blue curves). These curves yield pulses that cancel noise to second order in the case of a noisy level crossing. This collection of pulses can implement $z$-rotations of any angle $\phi$. The curve in Fig.~\ref{fig:phaseplots}a corresponds to a $\phi=\pi/4$ phase gate, while the curve in Fig.~\ref{fig:phaseplots2}a corresponds to a $\phi=\pi$ phase gate. Each curve in the family is comprised of a circle in the upper half plane and a smoothened parallelogram in the lower half. The corners of the parallelogram are made from circular segments. Because the rotation angle $\phi$ is determined by the angle between the initial and final tangent vectors~\cite{zeng2018general}, it is straightforward to adjust this angle by deforming the parallelogram to change the final tangent vector. A corresponding change in the area enclosed by the parallelogram can be compensated by resizing the circle, so that the net area enclosed by the entire curve remains zero. Since the curves consist of straight lines and circular arcs, the corresponding pulses are sequences of square pulses. The pulses corresponding to these curves are shown in Figs.~\ref{fig:phaseplots}b and \ref{fig:phaseplots2}b. The pulse heights and durations for each segment of the pulse in
Fig.~\ref{fig:phaseplots}b are 
\begin{align*}
    \Omega(t) T=
    \begin{cases}
    -12.8073 & 0\leq t< 0.490593T\\
    0        & 0.490593T \leq t<0.560359T \\
    20.2711 & 0.560359T\leq t<0.676594T \\
    0        & 0.676594T\leq t<0.725925T \\
    20.2711        & 0.725925T \leq t< 0.76467T \\
    0        & 0.76467T \leq t< 0.814001T \\
    20.2711        & 0.814001T\leq t<0.930231T \\ 
    0        & 0.930231T\leq t\leq T 
    \end{cases}
\end{align*}
Fig.~\ref{fig:phaseplots} also shows a naive constant pulse (orange) for comparison. This is given by
\begin{equation*}
    \Omega_c(t)\times T=-\pi/4.
\end{equation*}

The pulse heights and durations for each segment of the pulse in
Fig.~\ref{fig:phaseplots2}b are 
\begin{align*}
    \Omega(t)\times T=
    \begin{cases}
    -14.4221 & 0\leq t<0.435664T\\
    0        & 0.435664T \leq t<0.538169T\\
    14.4221  & 0.538169T \leq t< 0.683391T \\
    0        & 0.683391T\leq t<0.723423T \\
    28.8442        &0.723423T \leq t< 0.759728T \\
    0        & 0.759728T\leq t<0.782168T\\
    28.8442        & 0.782168T \leq t<0.854779T\\ 
    -14.4221        & 0.854779T\leq t\leq T
    \end{cases}
\end{align*}
Fig.~\ref{fig:phaseplots2} also shows a naive constant pulse (orange) for comparison. This is given by
\begin{equation*}
    \Omega_c(t)\times T=-\pi.
\end{equation*}

The second-order robustness of these pulses are confirmed in Figs.~\ref{fig:phaseplots}c,d and \ref{fig:phaseplots2}c,d, which show the LZ probability as a function of the noise parameter $\delta$ and noise strength $\sigma_\delta$. \textcolor{black}{Above, we retain six significant digits in the pulse parameters to ensure that the LZ probability plots capture the residual dephasing error rather than pulse parameter truncation errors.} The LZ probabilities for uncorrected square pulses that perform the same gates are also included for comparison.

\section{Monotonic sweeps cannot cancel noise}\label{app:Kneser}

Here, we prove that a monotonic LZ sweep function $\Omega(t)$ cannot cancel noise. We do this by proving that a plane curve with monotonic curvature cannot have a self-intersection. This essentially follows from a result from differential geometry known as the Tait-Kneser nesting theorem. 

To understand the Tait-Kneser theorem, we first need to introduce two general, related concepts associated with plane curves: {\it osculating circles} and the {\it evolute}. At each point along the curve $\mathbf{r}(t)$, one can define a circle whose radius $\rho(t)$ is equal to the inverse of the curvature $\kappa(t)$ at that point: $\rho(t)=1/\kappa(t)$. The center of this circle is located a distance $\rho(t)$ from $\mathbf{r}(t)$ in the direction of the normal vector $\mathbf{n}(t)$. Denoting the location of the center of the osculating circle by $\mathbf{e}(t)$, we therefore have
\begin{equation}\label{eq:evolute1}
    \mathbf{e}(t)=\mathbf{r}(t)+\rho(t)\mathbf{n}(t).
\end{equation}
The vector $\mathbf{e}(t)$ defines a curve called the evolute. It is comprised of the set of centers of the osculating circles of $\mathbf{r}(t)$.

\begin{figure}
    \centering
    \includegraphics[width=0.95\textwidth]{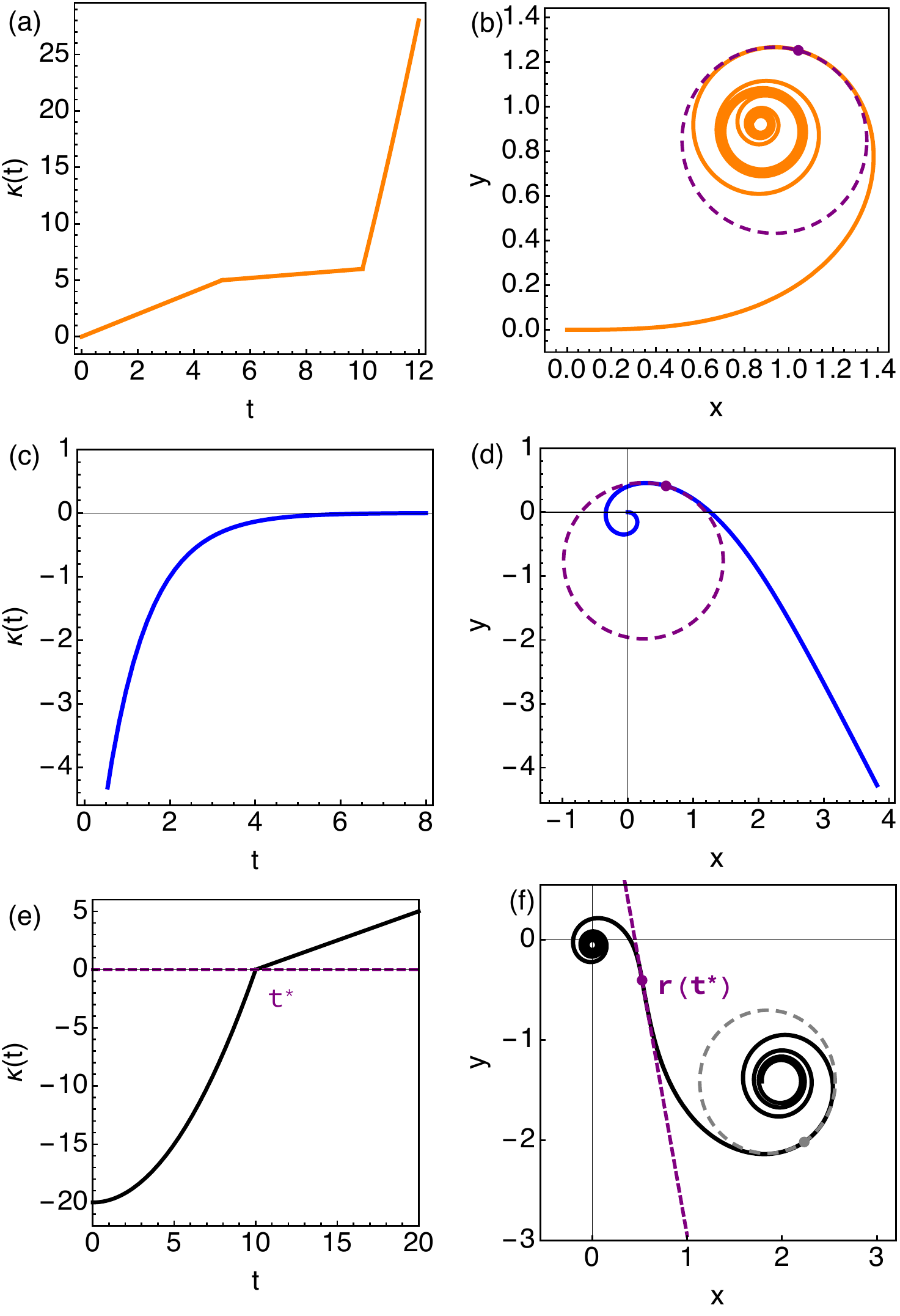}
    \caption{(a) An example of a non-negative, monotonically increasing curvature and (b) its corresponding curve. (c) An example of a non-positive, monotonically increasing curvature and (d) its corresponding curve. (e) An example of a monotonically increasing curvature that passes through zero and (f) its corresponding curve. The tangent line at the zero-curvature point $t^*$ is indicated with a blue dashed line. In all cases, no self-intersection occurs.}
    \label{fig:sweeptimes}
\end{figure}

The Tait-Kneser theorem states that if the curvature $\kappa(t)$ is a monotonically increasing function, then the osculating circles at later times are contained within the osculating circles at earlier times. Similarly, if $\kappa(t)$ is a monotonically decreasing function, then earlier osculating circles are contained within later ones. This can be proven in a few steps. First, observe that if the evolute does not contain any line segments, then
\begin{equation}\label{eq:evolute2}
    \|\mathbf{e}(t_1)-\mathbf{e}(t_0)\|<\int_{t_0}^{t_1}\|\dot{\mathbf{e}}(s)\|ds.
\end{equation}
Here, the right-hand side is the arc length of the evolute, and so the above inequality simply states that the arc length must be longer than the straight-line distance between the points $\mathbf{e}(t_1)$ and $\mathbf{e}(t_0)$. Next, notice that differentiating Eq.~\eqref{eq:evolute1} gives
\begin{equation}
    \dot{\mathbf{e}}=\dot{\mathbf{r}}+\dot\rho\mathbf{n}+\rho\dot{\mathbf{n}}=\dot\rho\mathbf{n},
\end{equation}
where we used the Frenet-Serret equation $\dot{\mathbf{n}}=-\kappa\dot{\mathbf{r}}$ to cancel two of the terms. We then see that $\|\dot{\mathbf{e}}\|=|\dot\rho|$. If $\kappa(t)$ is a monotonically increasing function, then $\dot\rho\le0$, and so $|\dot\rho|=-\dot\rho$. Plugging this into Eq.~\eqref{eq:evolute2}, we then obtain
\begin{equation}
    \|\mathbf{e}(t_1)-\mathbf{e}(t_0)\|<-\int_{t_0}^{t_1}\dot\rho(s)ds=\rho(t_0)-\rho(t_1).
\end{equation}
Thus, the osculating circle at point $t_1$ is fully contained inside the osculating circle at point $t_0$. If $\kappa(t)$ is instead a monotonically decreasing function, then $\dot\rho\ge0$ and $|\dot\rho|=\dot\rho$, which instead yields
\begin{equation}
    \|\mathbf{e}(t_1)-\mathbf{e}(t_0)\|<\int_{t_0}^{t_1}\dot\rho(s)ds=\rho(t_1)-\rho(t_0).
\end{equation}
In this case, the osculating circle at time $t_0$ is contained within the circle at $t_1$. This concludes the proof of the theorem.

An important corollary of the theorem is that a curve with monotonic curvature cannot have a self-intersection. This follows simply from that the fact that every point $\mathbf{r}(t)$ along the curve lies on its associated osculating circle of radius $\rho(t)$ centered at $\mathbf{e}(t)$. If the osculating circles are nested, as we just showed is the case for a monotonically changing curvature, then $\mathbf{r}(t_1)$ can never lie on the osculating circle associated with $\mathbf{r}(t_0)$, and so there cannot be an intersection. This corollary immediately implies that monotonic LZ sweeps cannot cancel noise: If the curve cannot self-intersect, then it is not possible to close the curve as necessary for first-order noise cancellation.

One slight subtlety with the above argument is that it implicitly assumes the curvature is always positive. This is potentially concerning for the LZ problem since $\Omega(t)$ can vanish during the sweep. If the curvature vanishes at some point $t^*$ along the curve, then the nesting theorem needs to be slightly modified. When the curvature vanishes, the radius diverges, and so at times just before $t^*$ the osculating circle becomes the entire half-plane lying on one side of the line parallel to the tangent vector $\dot{\mathbf{r}}(t^*)$ at that point. All osculating circles up to time $t^*$ will be contained (and nested within each other) within this first half-plane. At times just after $t^*$, the normal vector switches over to the other side of the tangent line at $t^*$, and the osculating circle now fills the second half-plane. All osculating circles at times later than $t^*$ are again nested and contained within this half-plane. The piece of the curve on either side of the tangent line at $t^*$ cannot intersect itself, nor can the two pieces intersect each other since they are confined to separate half-planes. These statements are illustrated in Fig.~\ref{fig:sweeptimes}. Therefore, no self-intersections can occur, and noise cancellation remains impossible for monotonic sweeps.


\section{Scaling of the Landau-Zener transition probability with LZ velocity}\label{app:scaling}
    
\begin{figure}[h]
    \centering
    \includegraphics[width=0.95\textwidth]{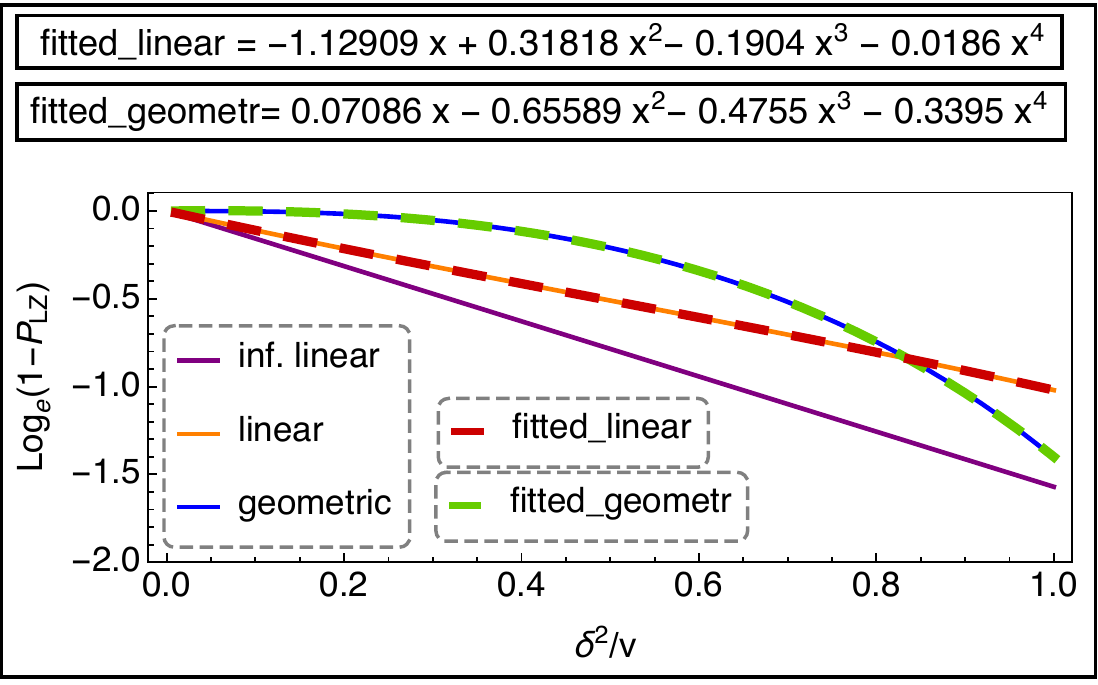}
    \caption{\textcolor{black}{Dependence of the Landau-Zener transition probability $P_{LZ}$ on $\delta^2/v$, where $\delta$ is the energy gap at the avoided crossing, and $v$ is the LZ velocity. Results are shown for an infinite linear sweep (purple), a finite-duration linear sweep (orange), and the geometrically engineered sweep from Fig.~\ref{fig:oddSweepLZAllPlots}f (blue). The red dashed and green dashed lines show quartic fits to the finite linear and geometrically engineered cases. The fitted parameter values are shown at the top of the figure.}
    }
    \label{fig:appendixCfig1}
\end{figure}

\textcolor{black}{As reviewed in the main text, the transition probability for an infinite, linear sweep is given by $P^{\infty}_{LZ}=1-e^{-\pi\delta^2/2v}$, where $\delta$ is the energy gap at the avoided crossing, and $v$ is the LZ velocity~\cite{landau1932,zener1932non}. The exponential dependence on $\delta^2/v$ is a hallmark of LZ physics. It is natural to ask how this dependence on $\delta$ and $v$ changes when a geometrically engineered pulse is used instead. To investigate this question, we compute the logarithm of $1-P_{LZ}$ as a function of $\delta^2/v$. For the original Landau-Zener formula, $P^{\infty}_{LZ}$, the result is a straight line with slope $-\pi/2$, as shown in Fig.~\ref{fig:appendixCfig1}. On the other hand, for the geometrically engineered pulse shown in Fig.~\ref{fig:oddSweepLZAllPlots}f, a qualitatively different behavior is evident, appearing quadratic at small $\delta^2/v$. To examine this behavior more closely, we fit $\log_e(1-P_{LZ})$ to a quartic polynomial in $\delta^2/v$. We find that the dominant term is indeed quadratic with coefficient -0.65589, indicating that the probability exhibits a Gaussian-like dependence on $\delta^2/v$ instead of an exponential one. The suppression of the linear term is a direct consequence of the error-cancellation property of the geometrically engineered pulse. At larger values of $\delta^2/v$, the transition probability also possesses significant cubic and quartic terms in the exponent. For comparison, the figure also shows the transition probability for the finite-duration linear sweep from Fig.~\ref{fig:oddSweepLZAllPlots}b. It is evident that truncating the sweep reduces the coefficient of the linear term in $\log_e(1-P_{LZ})$ from $-\pi/2$ to -1.12909, but the qualitative behavior with $\delta^2/v$ remains the same.
}

\end{document}